\documentclass[namedreferences]{SolarPhysics}
\usepackage[optionalrh]{spr-sola-addons} 
\usepackage{graphicx}                    
\usepackage{color}                       
\usepackage{url}                         

\begin{document}

\begin{article}

\begin{opening}

\title{THE BUTTERFLY DIAGRAM IN THE 18TH CENTURY}

%
\author{R.~\surname{Arlt}
       }

%

%
  \institute{$^{1}$ Astrophysikalisches Institut Potsdam,
                  An der Sternwarte~16, D-14482 Potsdam, Germany
                     email: \url{rarlt@aip.de}
             }

\begin{abstract}
Digitized images of the drawings by J.C.~Staudacher were used to
determine sunspot positions for the period of 1749--1796. From 
the entire set of drawings, 6285~sunspot positions were obtained 
for a total of 999~days. Various methods have been applied to find 
the orientation of the solar disk which is not given for the vast
majority of the drawings by Staudacher. Heliographic latitudes and
longitudes in the Carrington rotation frame were determined. The
resulting butterfly diagram shows a highly populated equator during
the first two cycles (Cycles~0 and~1 in the usual counting since
1749). An intermediate period is Cycle~2, whereas Cycles~3 and~4 show 
a typical butterfly shape. A tentative explanation may be the transient
dominance of a quadrupolar magnetic field during the first two cycles.
\end{abstract}

%

\end{opening}

%
\section{Introduction}
\label{Introduction} 
Sunspots typically appear at latitudes between $10^\circ$ and
$40^\circ$ heliographic latitude. At the beginning of the solar 
cycle, latitudes tend to be relatively high, while the appearance 
locations of sunspots shift to low latitudes as the cycle goes 
on. A plot of the spot appearance latitudes versus time
leads to the butterfly diagram (Maunder, 1904). The diagram
is typically plotted starting in 1874 with the Greenwich
drawings of the solar disk. The shape of the appearance
latitude as a function of time has not significantly changed
since. Regular updates of the butterfly diagram are provided
by Hathaway in publications and in the Internet (see e.g.
Hathaway {\it et al.}, 2003).

It is desirable  to extend the butterfly diagram into the
past. Especially after the Maunder minimum, the butterfly
diagram may tell us about the characteristics of the solar
dynamo when it was coming back to normal after an activity
lull with very few sunspots seen between 1645 and~1715.

There is a considerable set of drawings of the solar disk
made by Johann Staudacher from 1749 to~1796. The total set
of 848~drawings were digitized and described by Arlt (2008,
hereafter Paper~I). The drawings contain information about
1031~days in the above period, with several days combined
in one drawing, and notes about days when nothing was seen 
(the number has slightly increased since Paper~I, because
of additional notes taken into account).
For 999~of them, sunspots were plotted. Note that Wolf (1857)
was aware of Staudacher drawings and counted the sunspots
for his sunspot number time series which is still used today.
But the positions have never been determined since.

In this study, we present sunspot positions for 999~of the 
drawings by Staudacher and present the first butterfly diagram 
obtained for the 18th~century. Section~\ref{rotational_direction}
evaluates the direction of the rotation in the drawings which
appeared to have changed during the entire observing period of 
47~years. Section~\ref{orientation} deals with the various methods 
to derive the orientation of the solar equator, and Section~\ref{positions} 
describes the actual position measurements once the equator is
given. Section~\ref{butterfly} 
shows the results of the coordinate determinations in form of a 
butterfly diagram and discusses its features and limitations.
Finally, Section~\ref{discussion} discusses possible implications
for the theory of the solar dynamo.

\section{Rotational direction\label{rotational_direction}}
Even though the rotation is obvious in many sequences of drawings
over several days, there is still an ambiguity between a ``normal''
image and a mirrored, upside-down image. The two cases are geometrically
distinct only by the inclination of the solar rotation axis against
the ecliptic, which is far less obvious.

While the rotational direction was from right to left -- according to
a mirrored, upright projection image -- in all images until the end
of~1760, spots appear to move from left to right starting in~1761.
An obvious occasion is already the pair of observations of 1761 Feb~20
and Feb~24. If the images were still projected, north must then be 
at the lower border. The observation of 1761 May~25 has indeed ``S\"ud"
(south) at the  upper solar limb.

Did Staudacher stop using the mirrored image of the projection, or
did he turn the images by $180^\circ$? The eclipse of 1753 is an upright 
but mirrored image, just as the sunspot drawings. The eclipse of 1769 
is a mirrored image, turned by $180^\circ$ which is consistent with 
the upside-down sunspot drawings starting in 1761. The eclipse of 1791 
does not show the direction of the motion of the Moon, but the geometry 
and the indications of south pole and north pole, which are assumed to 
point to the celestial ones, are very consistent with a mirrored and 
rotated image, just as it was in 1769.

The assumption of rotated, but still mirrored images is backed up by 
further annotations with compass directions which are listed in 
Table~\ref{tab_orientations}. 
These are -- together with 
the lunar motion in solar eclipse drawings -- all the indications available,
and we assume that all images are mirrored throughout the period of 1749--1796.
Before any of the operations described in the following sections, all
images were mirrored. The change in orientation may mean that Staudacher 
used a Keplerian telescope until 1760 and a Gregorian starting in 1761, 
but this is speculation.

\begin{table}
\caption{Orientation indications given by Staudacher besides
solar eclipses.}
\label{tab_orientations}
\begin{tabular}{l@{~}l@{~}lllll}
\hline
\multicolumn{3}{l}{Date}        & North & East & South & West \\
\hline
1760&Sep&26 & top   & right&       &      \\
1761&May&25 &       &      & top   &      \\
1762&May&04 &       & left & top   &      \\
1762&May&05 &       & left & top   &      \\
1762&Nov&21 &bottom & left &       &      \\
1762&Nov&23 &bottom &      &       & right\\
1770&Jul&10 &       &    &top-right&      \\
1773&Aug&14 &bottom-left&top-left& &      \\
1773&Oct&17 &bottom-left&  &       &      \\
1774&Jun&16 &bottom-right&bottom-left & & \\
1774&Jul&10 &bottom-left&  &       &      \\
1775&Jun&08 &bottom-right& &       &      \\
1777&May&31 &bottom-right& &       &      \\
1785&Apr&17 &bottom-right& &       &      \\
\hline
\end{tabular}
\end{table}

\section{Position angles of the disks\label{orientation}}
There is no single way of fixing the orientation of the drawings.
We have to rely on various methods with various uncertainties.
An automatic method to obtain the sunspot positions is not applicable.
All the sunspot positions have been determined manually with a 
few IDL routines specially programmed for the Staudacher images, and 
the knowledge about how sunspot groups typically appear on the sun 
(today). The various ways of obtaining the solar equator are described 
in the following subsections. Four IDL programmes were written for 
the methods in Sections~\ref{rotation_fitting}, \ref{rotation_matching},
\ref{group_alignment}, and~\ref{time_of_day}.

\subsection{Rotation fitting\label{rotation_fitting}}
The method which is very frequently used in this study is the
rotational fitting of observations which are a few days apart.
If these show at least two common sunspot groups, the two disks
can be turned against each other, in order to achieve a
pattern motion consistent with the solar rotation. The more
sunspot groups are common to the drawings of both days, the
more reliable is the fit.

\begin{figure}    
\begin{center}
\includegraphics[width=\textwidth,clip=]{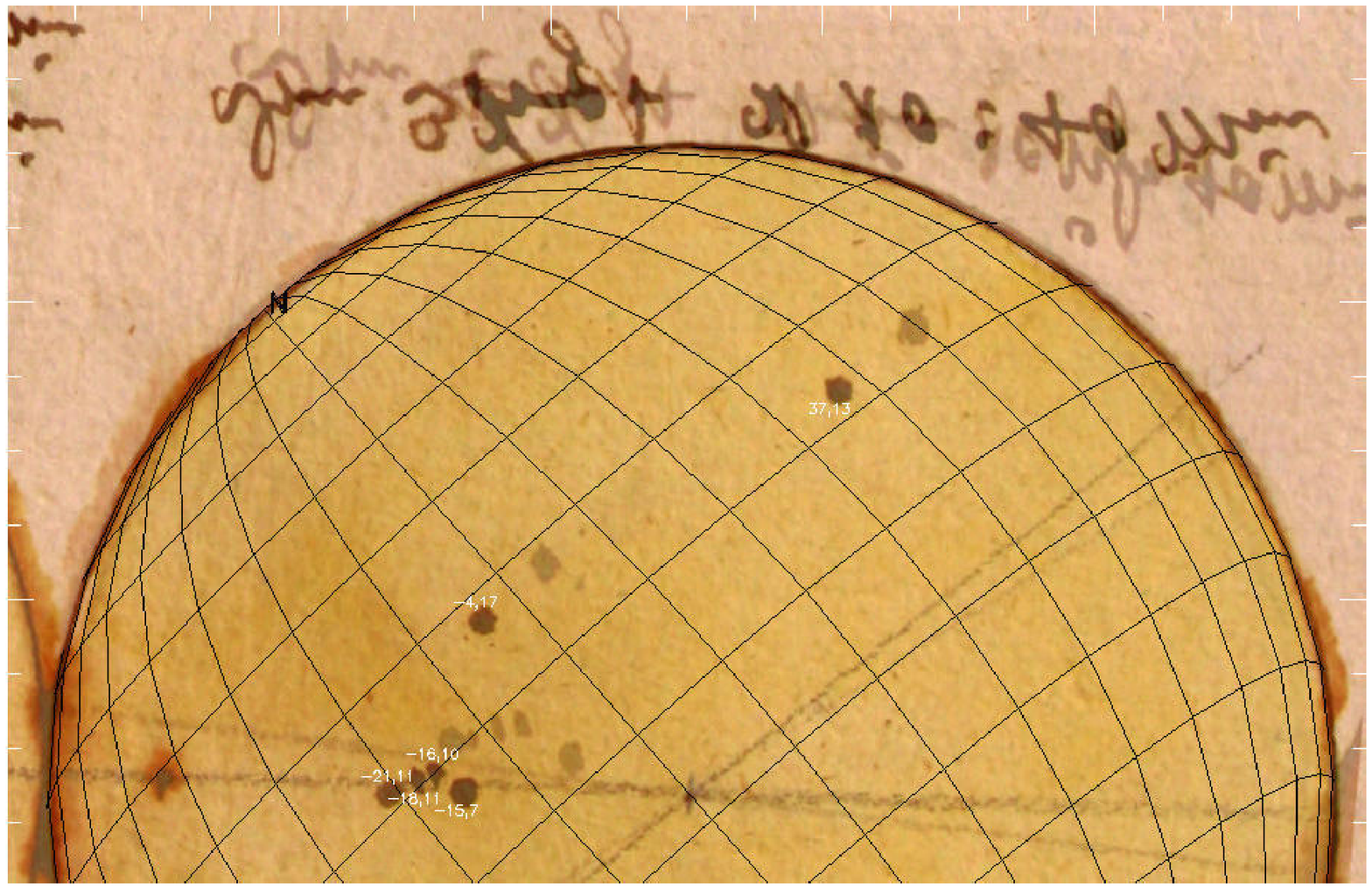}
\includegraphics[width=\textwidth,clip=]{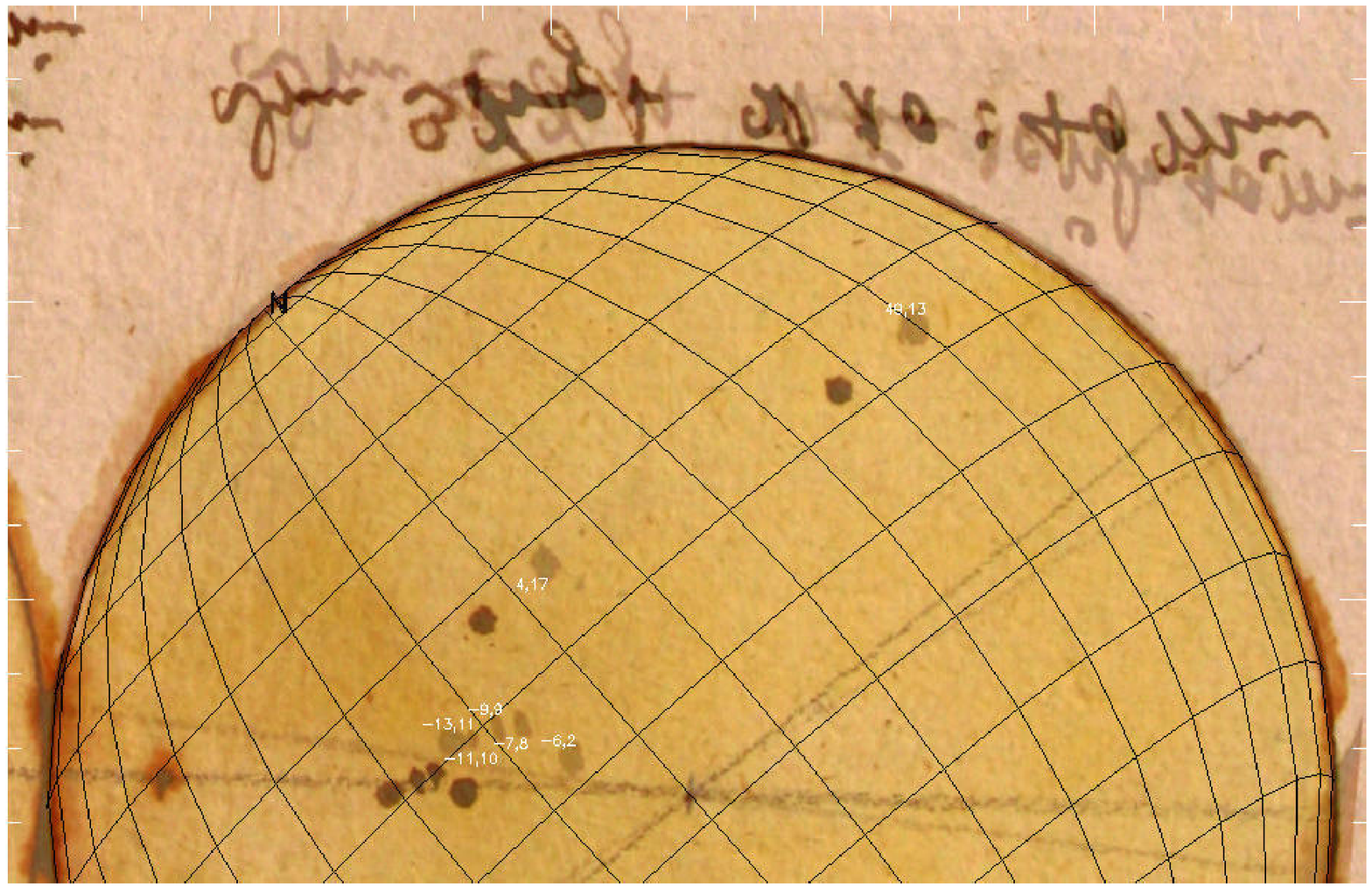}
\end{center}  
\caption{Rotation matching example of the two superimposed drawings of
1767 Sep~3 (dark spots in top panel) and~4 (grey spots in 
bottom panel). The images are rotated against each other, and 
scaled to have the same size.  Note that all images were mirrored
since they are the result of the projected Sun.}\label{rotation_fit}
\end{figure}

Due to the possible drawing errors by Staudacher, a fully automatic
fit has not been applied. Given two images with not more than about 
6~days time difference, one associates spots on one image with spots 
on the other. The optimum position angles for both drawings is determined 
numerically from these spot pairs. If more than one pair is available, 
there is, in general, a unique best fit. Figure~\ref{rotation_fit}
shows an example of superimposed drawings of two consecutive days.
The spots coming from 1767 Sep~03 are dark grey, the spots of 1767 Sep~04 
are shown in lighter grey. The second drawing is rotated in order to 
match the spot motion best.

We denote individual spot positions in the $n$-th drawing by Staudacher
with $L_i^{(n)}(\phi^{(n)}), B_i^{(n)}(\phi^{(n)})$ where $i$ runs over 
the spots drawn for a specific day. These are functions of the actual 
position angle $\phi^{(n)}$ chosen for the transformation of pixel 
coordinates into heliographic coordinates. The longitudes $L_i$ refer to the 
center-of-disk meridian until we get to the Carrington frame in
Subsection~\ref{carrington}. We use $n=1$ and $n=2$ referring to any 
given pair of drawings in the following. We need to find the optimum position
angles $\phi^{(1)}$, $\phi^{(2)}$ for any pair of images chosen.

The optimization requires a solar surface rotation profile, and 
we base the following procedure on the sideric rotation frequency
\begin{equation}
  \Omega_{\rm sid}(B) = 14.358 - 2.87 (\sin^2 B - \sin^2 15^\circ)
  \label{rotation_profile}
\end{equation}
in degrees per day, which was derived using sunspot positions, and where 
$B$ is the heliographic latitude (Balthasar, Vazquez, and W\"ohl, 1986). 
We approximate the synodic rotation frequency by $\Omega_{\rm syn} = 
\Omega_{\rm sid}-0.9867-0.0333 \cos(\lambda_\odot-283.4)$, where $\lambda_\odot$ 
is the solar longitude of the observing time. In the following, the synodic 
period $P=360^\circ/\Omega_{\rm syn}$ is used. 

A modified least squares fit delivers the best choice of the two position 
angles. For any given pair of $\phi^{(1)}$ and $\phi^{(2)}$, the positions
$(L_i^{(1)}(\phi^{(1)}), B_i^{(1)}(\phi^{(1)}))$ and 
$(L_i^{(2)}(\phi^{(2)}), B_i^{(2)}(\phi^{(2)}))$ are computed. In defining
a norm for the optimization, the latitude 
differences were chosen to enter the error sum with a fourth power rather than as
squares in order to give higher weight to the latitudinal fit as compared
with the longitudinal one. Since the rotation profile -- only known precisely 
from the last $\sim120$~yr -- may have been slightly different in the past, 
it is wise to give lower weight to the rotation period. The resulting position 
angles also looked subjectively better fitted than with equal powers. The 
square longitude differences are compared with the theoretically expected 
longitude shift $\Delta L_{\rm theor}$ according to (\ref{rotation_profile}): 
\begin{equation}
  \Delta L_{\rm theor} = \frac{2\cdot360^\circ\Delta t}{P(B_i^{(1)}) + P(B_i^{(2)})},
\end{equation}
where $\Delta t$ is the time difference between the two observations. The equation 
shows that the average rotation period according to the two different latitude 
representations is used. The function which is actually minimized is thus
\begin{equation}
  f(\phi_1,\phi_2)=
  \sum_i \left(B_i^{(1)}-B_i^{(2)}\right)^4 + 
  \cos^2 B_{\rm avg} \left(L_i^{(1)}-L_i^{(2)}-\Delta L_{\rm theor}\right)^2,
\end{equation}
All these coordinates $B$, $L$, and $L_{\rm theor}$ depend on the position 
angles $\phi^{(1)}$ and $\phi^{(2)}$ which are varied in order to find the 
best fit.  Since for arbitrary position angles, the corresponding latitudes
of the same spot in the two different drawings are not the same,
we use $B_{\rm avg}$ as the average of these two ``representations".

\subsection{Rotation matching\label{rotation_matching}}
There are quite a few occasions when there is at least a pair of 
dates for which a single spot is drawn. If these are directly stacked 
with transparency, one may have the impression that the equator is 
indicated by their connection line. But again, the position angles of 
the two drawings may differ (and they do severely). The rotation 
matching applied here is a special case of a rotation fit and delivers 
from zero to two exact solutions for the position angles.

Let us consider the position angle $\phi^{(1)}$ of the first
drawing. At a given $\phi^{(1)}$ there are generally two position
angles $\phi^{(2)}$ of the second drawing for which the spot
falls onto the same latitude as in the first drawing. The 
azimuthal difference between the longitude $L^{(1)}$ in the
first drawing and the longitude $L^{(2)}$ in the second drawing
plus the time difference between the two observations gives us
a measure for the corresponding rotation period.

The $\phi^{(2)}$ for which the ``rotation period'' is closest to the 
expected rotation period according to Balthasar, Vazquez, and W\"ohl 
(1986) is selected and delivers $P_{\rm match}(\phi^{(1)})$. The other 
match is discarded. This search of $\phi^{(2)}$ for a given $\phi^{(1)}$ 
is repeated for all possible $\phi^{(1)}=0$ to $2\pi$. The ``rotation 
periods'' obtained are compared with the true rotation period (assuming 
it has not changed significantly since the 18th century). A function 
$P_{\rm match}(\phi^{(1)})-P(B)$ is obtained, where $B$ is again 
the heliographic latitude (note that $B$ in both drawings are equal now). 
This can have a single minimum where there is 
an $P_{\rm match}(\phi^{(1)})$ getting close to $P$, or it can 
have two exact solutions where the period difference vanishes.  
If two solutions matching the rotation period of the corresponding
latitude exactly are found, the one for which sunspot latitudes fall
below $50^\circ$ was chosen. This also includes additional spots
in the drawings which have no counterpart in the other image.
This makes the final result a bit less objective, but the increase in
the number of available observations was considered more important.

\subsection{Direction lines}
A number of drawings show lines drawn by pencil apart from a
horizontal line often appearing to align images in rows in
the book. There is a total of 188~days for which such additional
lines are drawn; most of them are not annotated. Eight of them 
have markings referring to the ecliptic, 10 others have compass
directions as annotations (plus the first four drawings in 
Table~\ref{tab_orientations} with compass 
directions but no lines). These are too few to conclude anything
about the value for finding the orientations of the drawings;
some of them even contradict the distribution of the spots so
drastically, that we assume they are not meaningful for our
purposes. The lines might not even be inserted by Staudacher
himself actually.

\subsection{Alignment by sunspot groups\label{group_alignment}}
As long as solar activity is not at minimum, the orientation of
sunspot groups often provides a fairly accurate guess of the 
solar equator, especially if there are two or more groups. The
uncertainty essentially comes from the statistics of the tilt
angles of groups themselves. Howard (1996) determined the average
group tilt against the equator to be about $4^\circ$, withe the
preceding spots being closer to the equator than the following
ones. The scatter, however, is about $25^\circ$ around that
value. A single group thus provides limited orientation accuracy,
but a second group decreases the uncertainty a lot.

\subsection{Orientation by time of day\label{time_of_day}}
Finally, the drawings also indicate that Staudacher often placed his
drawing paper fairly well aligned with the horizon behind the
telescope. Times of day are given for observations starting in
1760. Starting with 1763, the orientations appear to be consistent
enough with an angle with the horizon. We start using this
indication of the drawing orientation occasionally -- if no other
way of determining the position angle was available -- in 1763 first.
There was no common reference time in Europe in the 18th century
yet, because of the lack of communication means. People in cities 
and villages most likely referred to a reference clock which was
adjusted according to astronomical events and thus showed something close
to local time. We assumed that Staudacher's clock times are local
time for Nuremberg.

Obtaining the drawing orientation from the time of day is the only method
where the geographical location of Staudacher matters. We assume a
geographical longitude of $\lambda=11.08^\circ$\,W and latitude of 
$\phi=49.45^\circ$\,N for the spherical transformations. The total
position angle of the solar rotation axis against the direction to the zenith
is determined from the tilt of the celestial equator against the
horizon, the tilt of the ecliptic against the equator, and the tilt 
of the solar equator against the ecliptic for the given date and time.
The orientations of 173~drawings were primarily based on the time
of day. For other drawings, if for example the spot distribution was
used primarily, the time of day helped as an additional support of 
other estimates made for the solar equator.

\subsection{Additional remarks}
The partial solar eclipse of 1769 June~4 was noted as observed on
June~3. This seemed to indicate that Staudacher used astronomical dates 
starting at noon (cf.\ Paper~I). However, the spot motion on consecutive
days has nowhere shown a discrepancy with dates starting at midnight.
An example is the pair of drawings on 1764 February~16, at 14h, and
1764 February~17, at 10h. While an astronomical day count would
imply that these two observations are 1.6~days apart, the spot
motion clearly shows that a difference of 0.8~days is much more likely.

In some cases, two position angles were about equally plausible.
The sunspot positions of both orientations were kept in the 
file of positions. The positions are stored in daily blocks separated
by blank lines. The results of two position angles are identified by 
two blocks with the same date (plus verbal remarks). Because of the 
limited accuracy of the drawings, we are seeking statistical results 
rather than results on individual spots. When these data are used, one 
should make sure a considerable set of positions is involved.

\section{Sunspot positions\label{positions}}
\subsection{Heliographic latitudes}
Once the orientation of the solar equator is defined, an ``observed''
heliographic coordinate system is established over the drawing. The
following operations have all been worked out with IDL's map routines
which are part of any recent IDL installation.
Note that the drawings are not necessarily precisely circular. Most of 
the distortions led to elliptical drawings with a longer vertical
axis and a shorter horizontal one. The ellipticity is taken into 
account by measuring both the horizontal and the vertical extent of the
circle drawn directly on the screen. Since IDL allows for non-circular
spherical grids, the setup of the surface map was straight-forward.

Setting up the coordinate system requires the determination of the
tilt of the sun against the observer. Because of the small inclination
of the sun's equator of $7.25^\circ$, we can use an approximation
for the annual variation of the tilt and apply
\begin{equation}
  B_0 = 7.25^\circ \cos (\lambda_\odot + 15^\circ)
\end{equation}
to obtain the latitude of the center of the solar disk, there $\lambda_\odot$
is the solar longitude for eq. J2000.0. The error compared with the much
more complex IAU2000 rotation model (Seidelmann {\it et al.}, 2002) is $0.35^\circ$
at maximum. The error in heliographic longitude is one order of magnitude
smaller. The heliographic positions $(L_{\rm obs},B)$ were then measured 
on the screen with the pixel accuracy of the digitized images which is
between $0.16^\circ$ and $0.26^\circ$ in the center of the solar disk
depending on the size of the photographic reproductions. Note that
the longitude $L_{\rm obs}$ still refers to the central meridian at the time of the
observation; the conversion into the Carrington frame is described in the
following Subsection.

A subjective quality tag $q$ is assigned to each day, essentially evaluating
the quality of fixing the equator. The ranking uses $q=1$ for very good,
$q=2$ for mediocre quality, and $q=3$ for unreliable positions. A rotational
fitting typically delivers $q=1$; in a number of cases, the fit was not
fully convincing and received $q=2$. Orientations obtained from group alignments
typically have $q=3$, if several groups were available or the
distribution in general supports a particular orientation, the quality was 
set to $q=2$. We have to keep in mind though that these position angles are
based on the knowledge of spot distribution and roughly horizontal alignment
of bipolar groups as they are observed {\it today}. These characteristics
may have been different at Staudacher's time. The position angles solely 
taken from the time of day received $q=3$.

\subsection{Heliographic longitudes\label{carrington}}
The longitudes of all the positions measured refer to the meridian
going through the center of the solar disk. The positions need to
be converted into the Carrington rotation frame in order to obtain
longitudes comparable to modern positional information. We need to
compute the heliographic longitude of the center of the solar disk
for any given time in 1749--1796. The method
described by Meeus (1985) is applied and was implemented in the 
IDL routines of the Applied Physics Laboratory at Johns Hopkins 
University ({\tt sun.pro} as of 1991 Jul~24).
The algorithm shows slight deviations from the IAU2000 rotation model
(Seidelmann {\it et al.}, 2002) of up to $0.33^\circ$ when going back to 1749. 
This is the model used by the HORIZONS ephemerides system at JPL with
which we compare here. The deviation is entirely linear, and we apply 
a correction
\begin{equation}
  L_{\rm 0,JPL} = L_{\rm 0,JHU} + 0.00111(2008-Y) + 0.041
\end{equation}
to the IDL routine, where $L_{\rm 0,JPL}$ and $L_{\rm 0,JHU}$ are 
the sub-observer heliographic longitudes of JPL and JHU respectively, 
and $Y$ is the year AD. The heliographic longitude $L$ in the Carrington
rotation frame is now simply $L = (L_{\rm obs} + L_{\rm 0,JPL}) \mathop{\rm mod} 360^\circ$. For
the observations without any note about the time of day (mostly before
mid-1761), local noon was assumed, corresponding to 11h16m~UT.

\section{The butterfly diagram\label{butterfly}}

\begin{figure}    
\begin{center}
\includegraphics[width=\textwidth,clip=]{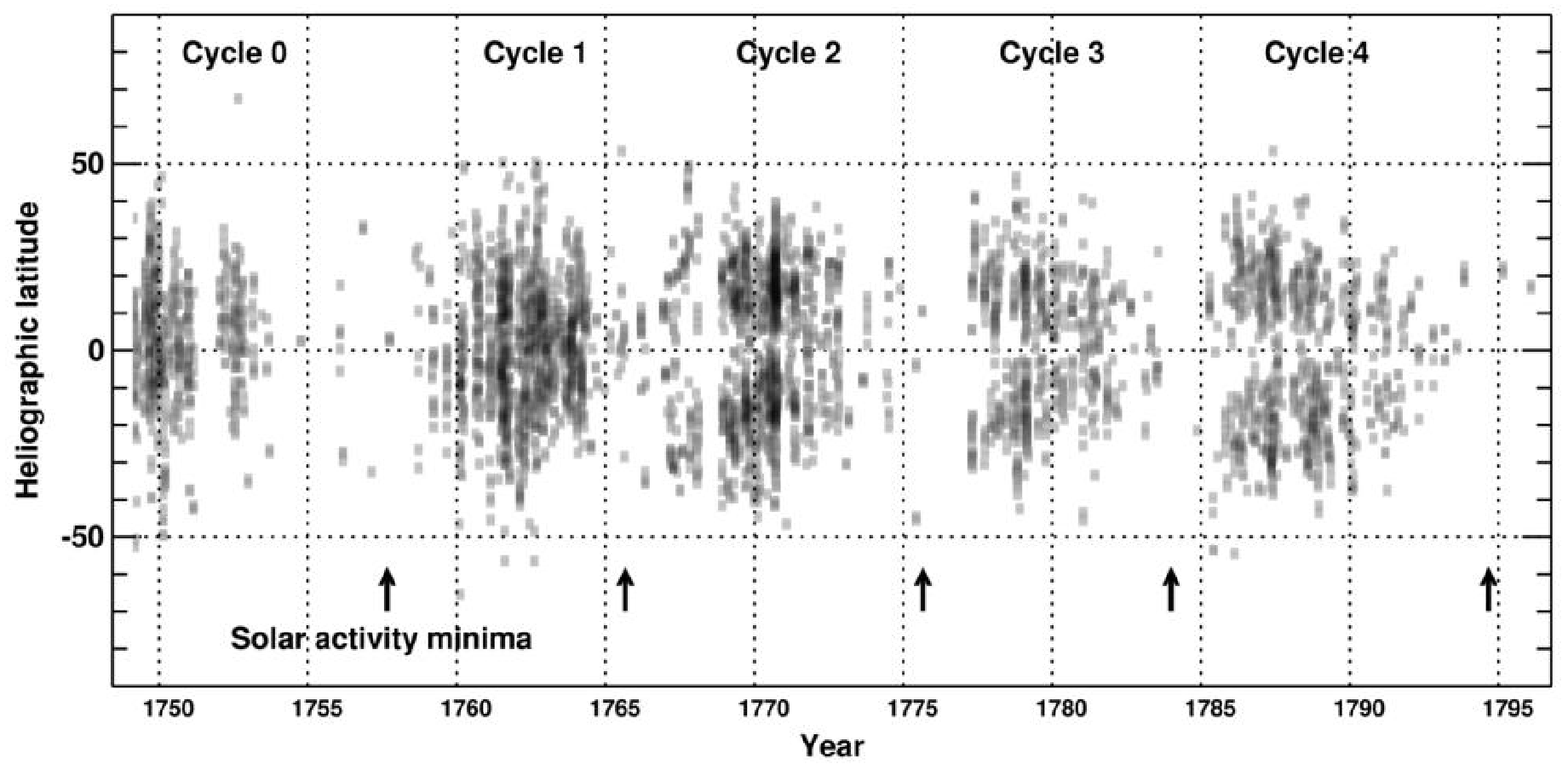}
\includegraphics[width=\textwidth,clip=]{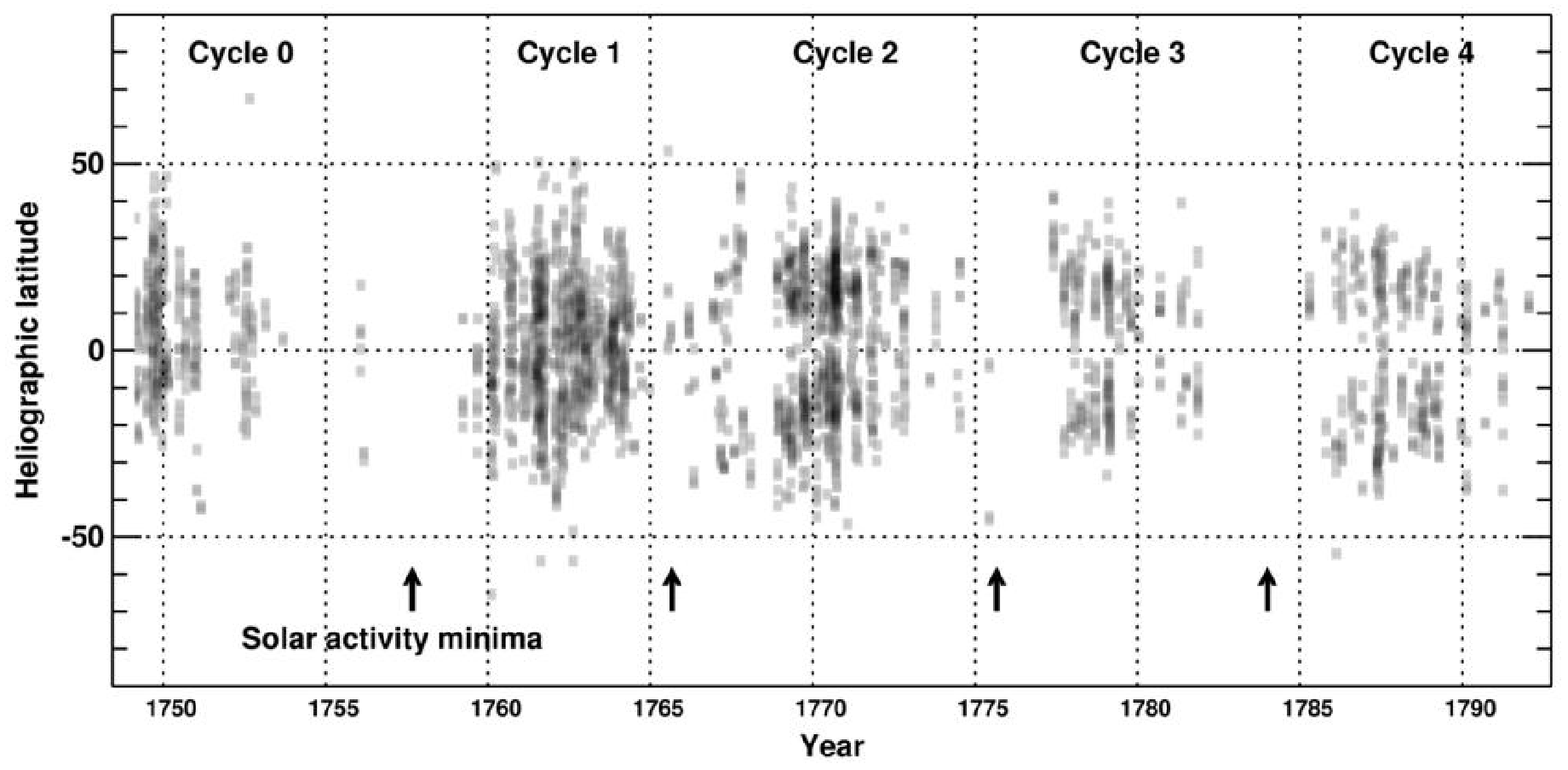}
\includegraphics[width=\textwidth,clip=]{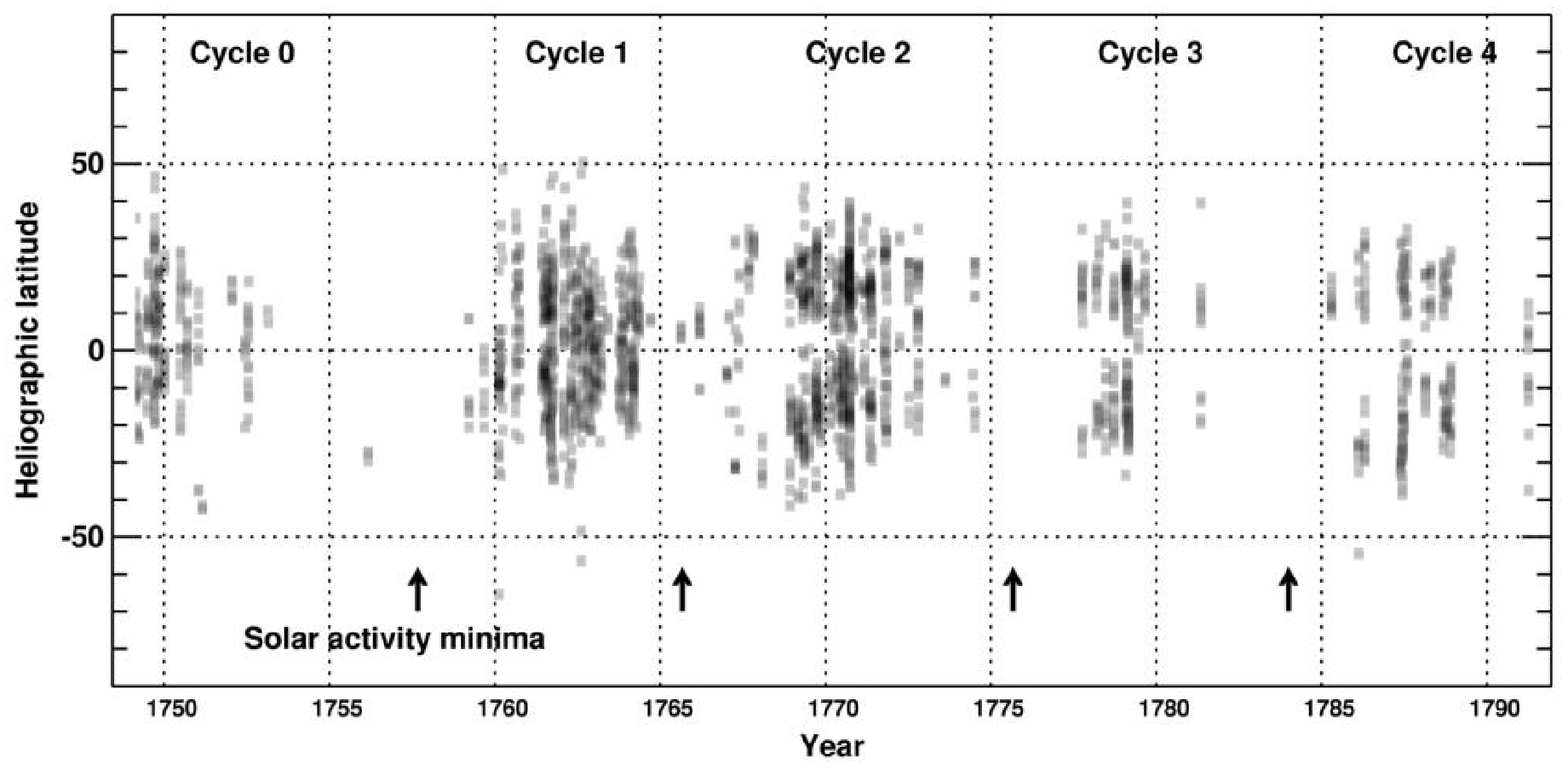}
\end{center}  
\caption{Butterfly diagram as derived from the drawings by Staudacher
in 1749--1796. The solar minima were taken from the sunspot area time
series of Paper~I. The top panels shows all spot positions, the middle
panel only spots with medium and high accuracy, the lower panel only
high accuracy.}\label{butterfly_diagram}
\end{figure}

The distribution of 6285 sunspot positions versus time and latitude is shown in 
Figure~\ref{butterfly_diagram}. The duration of spots is set artificially
to 50~days in order to increase the visibility of the data without altering the
results. Since in this first analysis, only the spot positions were measured and 
not the individual spot sizes, the latitudinal half-width of the spots is also 
artificial and was set to $\Delta B=2^\circ$. The $i$-th spot at latitude $B_i$
is distributed in latitude $b$ by a simple quadratic function
\begin{equation}
  d_i(t, b) = 1- \left(\frac{b-B_i}{\Delta B}\right)^2,
\end{equation}
where $d_i$ is a density function of time and latitude; it is set to 
zero outside $B_i \pm \Delta B$. The spots in
Figure~\ref{butterfly_diagram} thus have a total width of $4^\circ$.
All these spot densities are added for a single distribution $d(t,b) =\sum_i d_i(t, b)$.

The top panel shows the distribution of all positions measured, 
i.e. with a quality tag of $q\leq3$. The middle panel shows only
positions with $q\leq2$ while the lower panel shows $q=1$ spots
exclusively. The times marked with ``Solar activity minima" are the
times of lowest activity according to the sunspot area measurements
presented in Paper~I. The minimum times resulting from the Wolf
number series may be somewhat different.

The striking feature of Figure~\ref{butterfly_diagram} is the deviation
of the spot distribution from the butterfly shape during the period of
1749--1766. These are Cycles~0 and~1 in the cycle counting based on the
Wolf numbers. The solar equator is much more populated by sunspots than
we know it today. Also the migration toward lower latitudes is less
obvious, especially for Cycle~1 which is well covered by observations.
The following Cycle~2 also shows a populated equator but the butterfly
shape appears to be already in transition to the typical distribution 
which is then exhibited by Cycles~3 and~4.

The findings are apparently not altered when selecting only the more
reliable positions shown in the middle and lower panel of 
Figure~\ref{butterfly_diagram}.

\section{Discussion\label{discussion}}
Positional measurements of sunspots observed by Johann Staudacher in the period 
of 1749--1796 are presented. The data provide us with a butterfly diagram of almost 
five solar cycles covered. The distribution shows a surprising behaviour.
The first two cycles (Cycle~0 and~1) exhibit a different behaviour than
the one we know from the ``modern'' butterfly diagram as recorded since
1874. The highly populated equator indicates the presence of a dynamo
mode which is symmetric with respect to the equator. We do not have the
magnetic polarities though. Nevertheless, a polarity change across such
a populated equator seems very unlikely.

An interesting option is a differential rotation different from today.
The assumptions made for the rotational fits and matches will then
be less suitable. It is unlikely that the differential rotation changed
dramatically during the 18th century, but it will be an interesting
topic for future research to actually derive the rotation profile from
the spot drawings.

The results presented here may also be affected by a gradual change of 
observing skills and knowledge of Staudacher. The precise reproductions
of the solar eclipses -- with the first one drawn in 1753 -- indicate, 
however, that Staudacher projected the image directly on the paper he 
used for drawing. Such direct copies are likely to be fairly accurate at 
different stages of knowledge and experience. While there is certainly 
some development of skill in detecting small sunspots, the positional 
accuracy should remain fairly constant.

The sunspot cycle has not always shown sunspots as frequent as during the 
last 250~years. Strongly reduced activity was observed in the Maunder 
minimum period from about 1645 to 1715. The very few spots were 
then mostly present on the southern hemisphere (Ribes and Nesme-Ribes, 1993), 
apart from very few exceptions, until 1711. The cycle near the end of the 
Maunder minimum starting in about 1713 shows both hemispheres being populated 
again, with a considerable number of spots below $10^\circ$ heliographic 
latitude. Two activity cycles are missing before the results presented 
here take over. It was recently suggested that the north-south asymmetry 
may also be relevant for the prolonged Cycle~4 (Zolotova and Ponyavin, 2007). 

A number of dynamo models show the appearance of grand minima as a result 
of nonlinear effects between the magnetic fields and the differential rotation
(Tobias, 1997; K\"uker, Arlt, and R\"udiger, 1999; Bushby, 2006). Recent mean-field models 
with a meridional circulation controlling the solar cycle have been successful in 
constructing solar-like activity cycles (Dikpati and Charbonneau, 1999). The butterfly 
diagram is thought to be a direct consequence of the equator-ward flow at the 
bottom of the solar convection zone in these models. An interesting fact 
is the close excitation limits of the dipolar and quadrupolar dynamo modes 
in these flux-dominated mean-field dynamos (Dikpati and Gilman, 2001). A 
transient dominance of the quadrupolar (symmetric with respect to the equator) 
mode can either be explained by the chaotic nature of a deterministic system
(Weiss and Tobias, 2001) or by stochastic variations in the turbulence characteristics
(Brandenburg and Spiegel, 2008). Such variations have led to distortions of the 
butterfly diagram very similar to the ones observed by Staudacher in the 
18th century. The observations utilized here support the idea that the solar 
dynamo is modulated by nonlinear interactions between the dipolar mode -- 
dominant at the present time -- and a mode of quadrupolar symmetry.

Since the inspection and analysis requires a lot of man power, we
have derived here only the positions of the individual spots, but
have no information on the sizes of individual spots. There is very
likely information still hidden in the individual areas, but we
have to postpone these measuring efforts to a future project. The 
sunspot positions of 1749--1796 are available on request from the
author.

%

%

%

%
\begin{acks}
The author is grateful to Jan Meyer for the photographic reproductions
of the drawings and to Regina von Berlepsch for the support
in the library of the Astrophysikalisches Institut Potsdam. The work also 
benefited from the support by ISSI, Switzerland, where the utilization of
the drawings was discussed in an International Team.
\end{acks}

%
%
%

\end{article} 
\end{document}